\documentclass[fp,twocolumn]{jpsj3}
\usepackage{txfonts}
\usepackage{graphicx}
\usepackage{bm}
\usepackage{braket}
\usepackage{mathtools}
\usepackage{color}
\usepackage{ulem}
\usepackage{multicol}

\title{Orbital Magnetism in Honeycomb Ladder}

\author{Tomonari Mizoguchi\thanks{mizoguchi@rhodia.ph.tsukuba.ac.jp}$^1$,
Soshun Ozaki$^2$, 
and Hiroyasu Matsuura$^3$
}
\inst{$^1$Department of Physics, University of Tsukuba, 1-1-1 Tennodai, Tsukuba, Ibaraki 305-8571, Japan \\
$^2$Department of Basic Science, University of Tokyo, 3-8-1 Komaba, Meguro-ku, Tokyo 153-0041, Japan \\
$^3$Department of Physics, University of Tokyo, 7-3-1 Hongo, Bunkyo-ku, Tokyo 113-0033, Japan
}

\abst{
We investigate the orbital magnetic susceptibility 
of the tight-binding model for the honeycomb ladder
with the additional vertical hopping. 
Despite being one-dimensional, the magnetic flux penetrating the hexagonal rings affects the energetics of electrons, resulting in the finite orbital magnetic response.
We find that 
the orbital magnetic susceptibility is sensitive to the parameters as well as the chemical potential.
At half-filling, the response is diamagnetic when the system is close to the pure honeycomb ladder, whereas it turns to paramagnetic when it is close to the two-leg ladder. 
We also find several characteristic properties away from half-filling, such as the diamagnetic response at the band top and bottom, and the large paramagnetic response at the band gap sandwiched by the divergent density of states.   
}

\begin{document}
\maketitle

\section{Introduction}
Orbital magnetic responses of nonmagnetic solids provides us with the important information of their electronic structures. 
In the strong-field regime where Landau quantization occurs,
oscillatory behaviors of the physical quantities such as 
the de Haas-van Alphen effect 
and the Shubnikov-de Haas effect
are utilized to analyze the Fermi surface shape of metals~\cite{Abrikosov_text}.
In the weak-field regime, the orbital motion induced by the magnetic field contributes to the magnetic susceptibility~\cite{Landau1930,Peierls1933}.
When the spin-orbit coupling is weak,
such an orbital part contributes to the magnetic susceptibility independently to a spin part.
In contrast to 
the spin susceptibility which is determined solely by the density of states (DOS) 
at the Fermi surface and is ubiquitously paramagnetic,  
the orbital magnetic susceptibility has a much more complicated mechanism and can be either paramagnetic or diamagnetic.
In particular, numerous studies have revealed 
the importance of the interband effects~\cite{Fukuyama1971,Ogata2015},
and such effects are in close relation with the topology and geometry of the Bloch wavefunctions~\cite{Raoux2015,Gao2015,Ogata2015,Piechon2016,Kariyado2021, Murakami2006, NakaiNomura2015, OzakiOgata2021}. 

The naive intuition of the magnetic field effect on the orbital motion of the electrons are given by the cyclotron motion in the classical picture, or the wave packet dynamics on the equi-energy surface in the semi-classical picture~\cite{Xiao2010}.
Hence, most of the previous studies on the orbital magnetic susceptibilities are on the two or three dimensional systems.
Indeed, various formulas based on the Green's function method for calculating the orbital magnetic susceptibilities have been derived, but they mainly target higher-dimensional systems as they formulas containing the velocity operator in $x$ and $y$ directions~\cite{Fukuyama1971,Koshino2007,GomezSantos2011,Raoux2015,Gao2015,Ogata2015,Piechon2016,Kariyado2021}.  
In contrast, little has been studied 
on the orbital magnetic susceptibility
of one-dimensional models. 
If the lattice contains a plaquette-like structure (e.g., the two-leg ladder),
the magnetic flux penetrating the plaquette affects the orbital motion of electrons, resulting in the finite orbital magnetic response~\cite{Narozhny2005,Roux2007,Matsuura2016}.

\begin{figure}[b]
\begin{center}
\includegraphics[clip,width = \linewidth]{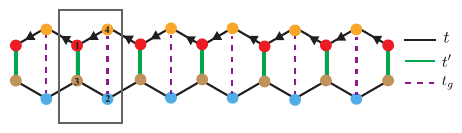}
\vspace{-10pt}
\caption{(Color online) Schematic figure
of the tight-binding model considered in this paper. 
The black solid, green solid, and purple dashed bonds 
represent the hoppings $t$, $t^\prime$, and $t_g$, respectively.
The bonds with an arrow 
have a phase factor $e^{i\phi/2}$.
}
\label{fig:model}
\end{center}
\vspace{-10pt}
\end{figure}

In this paper, we investigate the orbital magnetic susceptibility for the honeycomb-ladder tight-binding model with the additional vertical hoppings, shown in Fig.~\ref{fig:model}.
Deriving the exact single-particle eigenenergies and 
its flux derivatives, 
we calculate the orbital magnetic susceptibility without relying on the Green's-function-based formulas 
that treat the magnetic field effect perturbatively. 
(Nevertheless, as we will show, 
the results obtained with our method agree with 
those derived from the Green's-function-based approach.)
We reveal that response is highly dependent on the hopping parameters as well as the chemical potential, 
varying from paramagnetic to diamagnetic. 
At half-filling, we find that the sign of the orbital magnetic susceptibility is determined by the ratio of the parameters;
when the system is close to the pure honeycomb ladder, the orbital magnetic susceptibility is diamagnetic, whereas when it is close to the conventional two-leg ladder, the susceptibility turns to paramagnetic.
Several notable features are also found away from half-filling, 
such as the diamagnetic response at the band top and bottom, and the large paramagnetic response at the band gap sandwiched by the divergent density of states.
\begin{figure*}[tb]
\begin{center}
\includegraphics[clip,width = \linewidth]{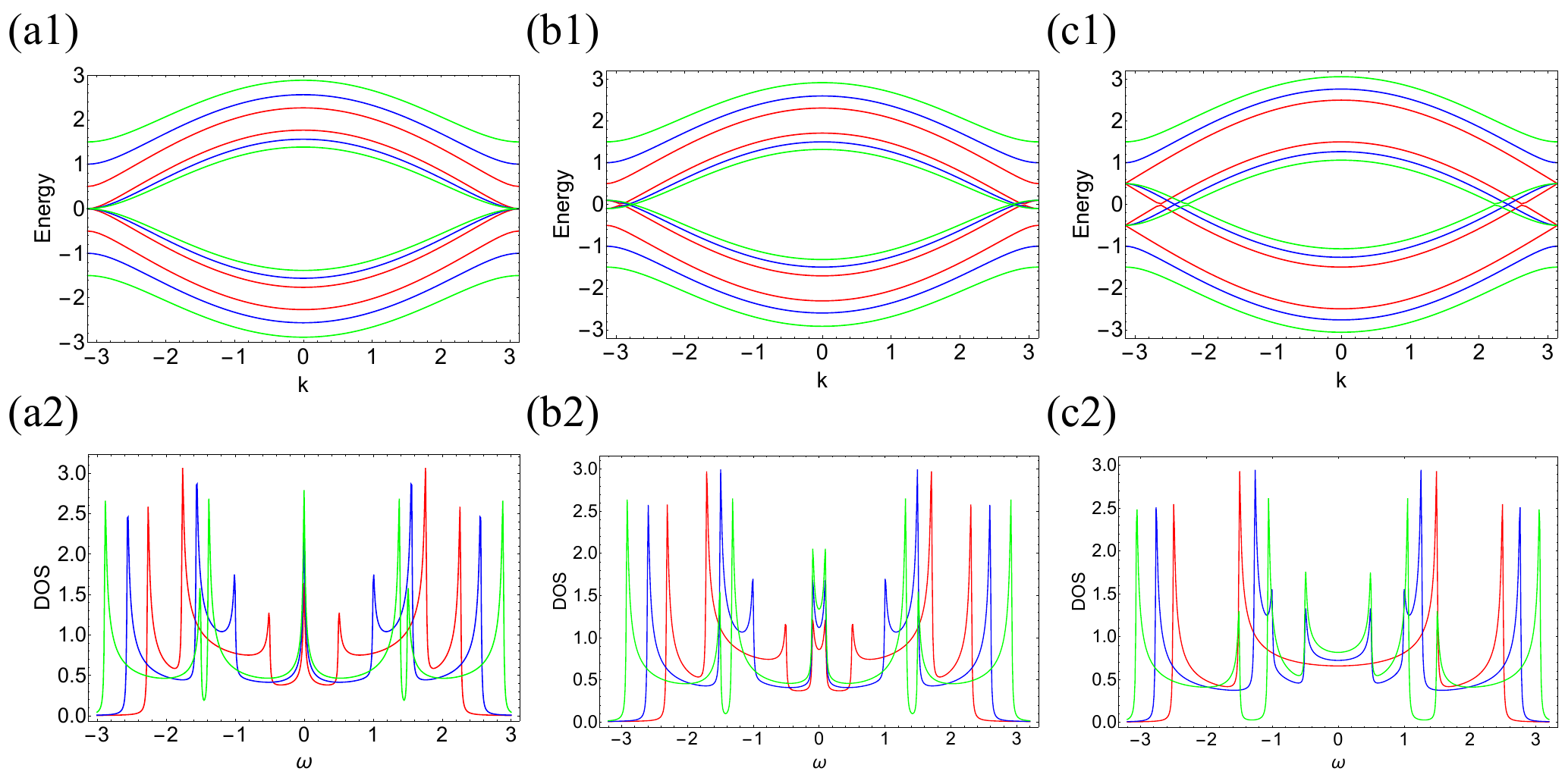}
\vspace{-10pt}
\caption{(Color online)
The band structure (upper panels) and DOS (lower panels)
for 
(a) $t_g = 0$
(b) $t_g = -0.1$
and (c) $t_g = -0.5$.
Red, blue, and greens lines
are for $t^\prime = -0.5$,
$t^\prime = -1$,
and $t^\prime = -1.5$, respectively.
We set $t=-1$ and $\phi = 0$.
}
\label{fig:band}
\end{center}
\vspace{-10pt}
\end{figure*}

We note that the model considered here can be regarded as an 
effective model for the polymer composed of C$_6$ rings, which is dubbed polyacene.
The model has been 
investigated as a 
one-dimensional correlated electron systems
and various exotic phenomena such as superconductivity and magnetism have been pursued~\cite{Kivelson1983,Karakonstantakis2013,Schmitteckert2017}.
From the experimental side, the length of the polyacene 
was~\cite{Liang2011,Watanabe2012,Christina2021}, but very recently  
the polyacene containing numerous number of C$_6$ rings was successfully synthesized~\cite{Kitao2023}.
Although it is difficult to extract the orbital magnetic susceptibility experimentally by separating it from the spin part, 
we believe that our result will contribute to understanding the fundamental properties of polyacene.

The model can also be regarded as 
the thinnest limit of the graphene nano-ribbon with the zigzag edge, when the additional vertical hoppings are absent. 
In this regard, Wakabayashi \textit{et al.,} investigated the magnetic field response of the graphene 
nano-ribbon 
with the zigzag edge with the generic width~\cite{Wakabayashi1999}, focusing on the half-filled case. 
In this work, on the other hand, 
we investigate the $\mu$ dependence of the orbital magnetic susceptibility (i.e., away from half-filling), as well as the nontrivial effect of the additional vertical hoppings ($t_g$).

The rest of this paper is organized as follows.
We first introduce the honeycomb ladder model and 
elaborate its characteristic band structure in Sect.~\ref{sec:model}.
In Sect.~\ref{sec:formulation}, we present the formulation of the calculation of the orbital magnetization and the magnetic susceptibility.
Our main results are presented in Sect.~\ref{sec:result}, where we argue the parameter dependence of the orbital magnetic susceptibility
at half-filling, the chemical potential dependence, 
and the real-space current distribution.
Finally, we present the summary of this paper in Sect.~\ref{sec:summary}.

In what follow, $\hbar$ stands for the reduced Planck constant,
$e(>0)$ stands for the elementary charge, 
and $k_{\rm B}$ stands for the Boltzmann constant.

\section{Model \label{sec:model}} 
We consider the tight-binding Hamiltonian depicted in Fig.~\ref{fig:model}~\cite{Kivelson1983,Karakonstantakis2013,Schmitteckert2017}.
The lattice contains four sublattices per unit cell,
denoted by 1-4. 
Hence each site is specified by the unit-cell position $R$ and the sublattice index $\alpha$ ($\alpha = 1,\cdots, 4$).
We describe the coordinate of the sublattice $\alpha$ measured from the origin of the unit cell as $r_\alpha$, which we will used later.
Throughout this paper, 
we neglect the spin degrees of freedom. 
We consider the system containing 
$L$ unit cells and impose the periodic boundary condition.
The length of the unit cell is set to be unity,
and we write the area of the hexagonal plaquette $S$. 
The effect of the magnetic flux  ($\phi$) is incorporated by the Peierls phase, denoted by the black arrows in Fig.~\ref{fig:model}.
The ``beyond-Peierls-phase" effect of the magnetic field~\cite{Matsuura2016_Pe} is not taken into account
in the present study.
\begin{figure*}[tb]
\begin{center}
\includegraphics[clip,width = \linewidth]{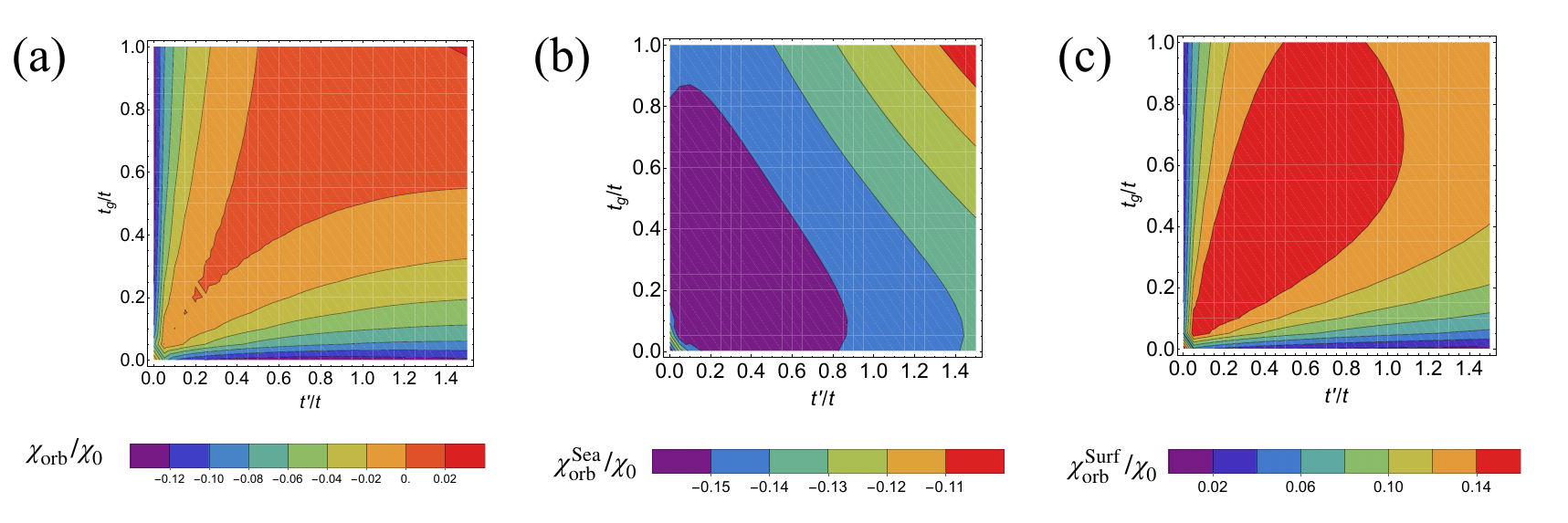}
\vspace{-10pt}
\caption{(Color online) 
Map of (a) $\chi_{\rm orb}/\chi_0$,
(b) $\chi^{\rm Sea}_{\rm orb}/\chi_0$,
and (c) $\chi_{\rm orb}^{\rm Surf}/\chi_0$
in the parameter space
$(t^{\prime} /t)$-$(t_g/t)$
at $t=-1$, $k_{\rm B}T=0.01$, and $\mu = 0$.
}
\label{fig:result_HF}
\end{center}
\vspace{-10pt}
\end{figure*}

Performing the Fourier transform, 
we obtain the momentum-space representation of the Hamiltonian, which reads
\begin{align}
H = \sum_{k} \bm{\psi}_{k}^\dagger \mathcal{H}_k (\phi) \bm{\psi}_{k},
\end{align} 
with $\bm{\psi}_k = \left( c_{k,1}, c_{k,2},c_{k,3},c_{k,4} \right)^{\rm T}$ 
are the column vector aligning the annihilation operators, 
and 
\begin{align}
\mathcal{H}_k (\phi)  = 
\begin{pmatrix}
\mathcal{O}_{2,2} & W^{\rm T}_k(\phi) \\
W_k (\phi) &  \mathcal{O}_{2,2} \\
\end{pmatrix}, \label{eq:Ham}
\end{align}
with
\begin{align}
W_k (\phi )
 = \begin{pmatrix}
 t^\prime & g_2(k) \\
 g_1(k,\phi) & t_g \\
 \end{pmatrix}.
\end{align}
Here we have defined
\begin{subequations}
\begin{align}
g_1 (k,\phi) = 2t \cos \left(\frac{k-\phi}{2}\right),
\end{align}
and 
\begin{align}
g_2 (k) = 2 t \cos \frac{k}{2}.
\end{align}
\end{subequations}
For future use, we also define
\begin{subequations}
\begin{align}
a(k,\phi) = t^{\prime 2} + [g_1 (k,\phi)]^2,
\end{align}
\begin{align}
b(k) =t_g g_1(k)+ t^\prime g_2 (k),
\end{align}
and
\begin{align}
c(k) = t_g^2 + [g_2(k)]^2.
\end{align}
\end{subequations}
Note that the Hamiltonian of Eq.~(\ref{eq:Ham}) preserves the chiral symmetry, namely, 
$\{\mathcal{H}_k(\phi), \gamma \} = 0$ holds 
with $\gamma = \mathrm{diag}(1,1,-1,-1)$.
Note also that, throughout this paper,
we set $t=-1$ and adopt $|t| = 1$ as a unit of the energy.

Solving the eigenvalue problem for $\mathcal{H}_k(\phi)$,
we obtain the eigenenergies which are given as 
\begin{align}
E =\varepsilon_{k;\eta_1, \eta_2}(\phi) = \eta_1 \sqrt{\Lambda_{k;\eta_2}(\phi)} \hspace{1pt} (\eta_{i} = \pm1), 
\end{align}
where
\begin{align}
\Lambda_{k;\eta_2}(\phi)
= \frac{a(k,\phi)+c(k)}{2}
+ \eta_2 \sqrt{\left(\frac{a(k,\phi)-c(k)}{2}\right)^2 + [b(k)]^2}.
\end{align} 
For later use, we introduce $\bm{u}_{k;\eta_1,\eta_2}(\phi)$
which denotes the eigenvector of $\mathcal{H}_{k}(\phi)$ 
with the eigenvalue $\varepsilon_{k;\eta_1, \eta_2}(\phi)$, 
and $u^\alpha_{k;\eta_1,\eta_2}(\phi)$ denotes the $\alpha$-th component of $\bm{u}_{k;\eta_1,\eta_2}(\phi)$.

Figure~\ref{fig:band} shows 
the band structure and DOS for $\phi = 0$.
Note that the DOS is given as 
\begin{align}
\mathrm{DOS}(\omega) 
= \sum_{\eta_1, \eta_2 = \pm1} 
\int_{-\pi}^{\pi} \frac{dk}{2\pi^2} \frac{\delta}{[(\omega-\varepsilon_{k;\eta_1,\eta_2}(\phi))^2+\delta^2]},
\end{align}
with $\delta$ being a small parameter, set as $\delta = 0.01$.
For all panels, we see that the positive and negative energy bands appear in a pairwise manner, which results from the chiral symmetry. 
For $t_g = 0$, 
the quadratic band touching occurs at $E=0$, $k=\pi$. 
Away from the zero-energy, we see several peak of the DOS, 
which arises from the band top or the bottom having the quadratic dispersion. 
It is noteworthy that, 
for $t_g =0$, the quadratic band touching 
at $E=0$ gives the divergent DOS.
For the finite $t_g$ (with $t^{\prime} t_g > 0$),
The quadratic band touching splits into two Dirac points.
For $\phi = 0$, the Dirac points are obtained as 
\begin{align}
k^{\mathrm{DP}} = \pm 2 \cos^{-1}\left(\frac{\sqrt{t^\prime t_g}}{2|t|}\right),
\end{align}
with the range of $\cos^{-1}$ is $[0,\pi]$~\cite{Kivelson1983}.

It is also worth mentioning that, for $t^\prime = t_g$, the system is equivalent to the conventional two-leg ladder~\cite{Hatsugai2006} whose orbital magnetism was investigated in Ref.~\cite{Roux2007}.

\section{Formulation \label{sec:formulation}}
In the present model, magnetization and magnetic susceptibility can be calculated by using standard statistical mechanics for grand canonical ensembles.
Namely, for the non-interacting 
fermions obeying the grand canonical distribution at the temperature $T$,
the thermodynamic potential is given as
\begin{align}
\Omega(\phi, \mu, T) = 
-\frac{1}{\beta} \sum_{\nu}
\ln \left(1 + e^{-\beta(\varepsilon_\nu (\phi) -\mu)}
\right), 
\end{align}
where
$\nu$ is the label of the single-particle state [in the present model $\nu =(k,\eta_1,\eta_2)$]
and $\beta = \frac{1}{k_{\rm B} T}$.
Then the orbital magnetization $M$ and the orbital magnetic susceptibility $\chi_{\rm orb}$
are, respectively, given as
\begin{align}
M = -\frac{\partial \Omega}
{\partial B}
= - A_0\frac{\partial \Omega}
{\partial \phi}
= -A_0 \sum_\nu \frac{\partial \varepsilon_\nu (\phi)}{\partial \phi} f(\varepsilon_\nu (\phi)), \label{eq:mag}
\end{align}
and 
\begin{align}
\chi_{\rm orb} = \frac{\partial M}{\partial B}
= -A_0^2 \sum_{\nu}
\left[ 
\frac{\partial^2 \varepsilon_\nu(\phi)}{\partial \phi^2} f(\varepsilon_\nu (\phi))
+ \left( \frac{\partial \varepsilon_\nu (\phi)}{\partial \phi}\right)^2 f^\prime(\varepsilon_\nu (\phi))
\right], \label{eq:chi}
\end{align}
where 
$B$ is the magnetic field which is related to the magnetic flux as
$\phi = A_0 B$ with $A_0
:= -\frac{e}{\hbar}S$, and 
$f(x)= \frac{1}{e^{\beta(x-\mu)} + 1}$ with $\mu$ being the chemical potential.
Clearly, the term proportional to $f(\varepsilon)$ represents the contribution comes from the Fermi sea, 
whereas that proportional to  $f^\prime (\varepsilon)$ from the Fermi surface.
Thus we call the former the Fermi sea term and the latter the Fermi surface term, which we write $\chi_{\rm orb}^{\rm Sea}$ and $\chi_{\rm orb}^{\rm Surf}$, respectively. 
It is worth noting that $\chi_{\rm orb}^{\rm Surf}$ is non-negative.

As we have the analytic form of the single-particle eigenenergies $\varepsilon_{k;\eta_1,\eta_2}(\phi)$,
the derivative of these with respect to $\phi$
can also be obtained straightforwardly.
In the following, we show the results obtained by performing the summation over $k$ numerically. 
\begin{figure*}[tb]
\begin{center}
\includegraphics[clip,width = \linewidth]{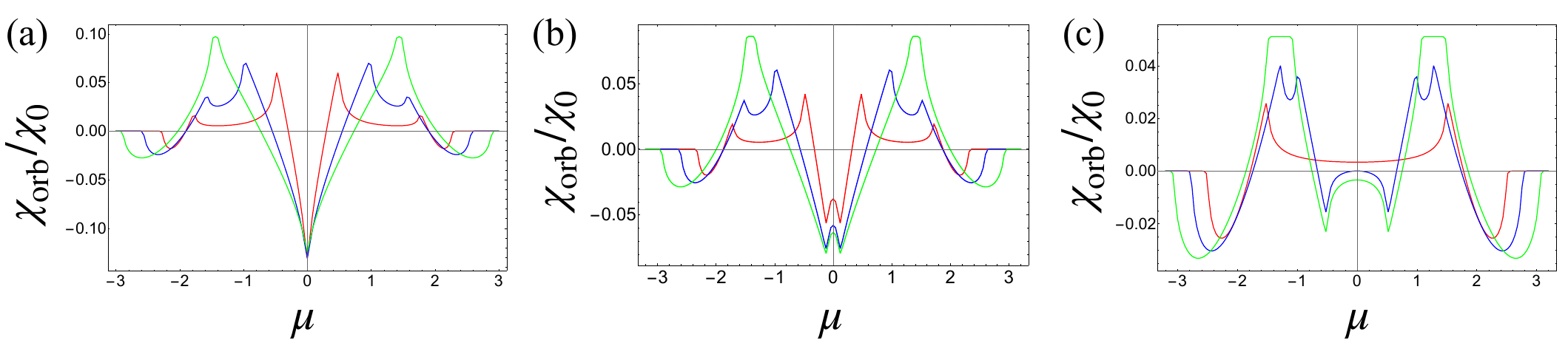}
\vspace{-10pt}
\caption{(Color online)
$\mu$ dependence of $\chi_{\rm orb}$ at $t=-1$ and $k_{\rm B}T=0.01$ with
(a) $t_g = 0$
(b) $t_g = -0.1$
and (c) $t_g = -0.5$.
Red, blue, and greens lines
are for $t^\prime = -0.5$,
$t^\prime = -1$,
and $t^\prime = -1.5$, respectively.
}
\label{fig:result}
\end{center}
\vspace{-10pt}
\end{figure*}

By using Eq.~(\ref{eq:chi}), we can compute the magnetic susceptibility for arbitrary $\phi$ and $T$.
In what follows, we focus on the magnetic susceptibility at $\phi = 0$ and $k_{\rm B}T=0.01$. 
We adopt $\chi_0 = A_0^2 L$ as a unit of the susceptibility.

Before proceeding further, 
we note that, in the present model, 
$M$ of Eq.~(\ref{eq:mag})
and $\chi_{\rm orb}$ of Eq.~(\ref{eq:chi})
satisfy what is called the sum rule with respect to $\mu$~\cite{GomezSantos2011,Gutierrez2016}, that is,
\begin{align}
 \int_{-\infty}^\infty M d\mu =0,
\end{align}
and 
\begin{align}
 \int_{-\infty}^\infty \chi_{\rm orb} d\mu =0,
\end{align}
hold for any $\phi$ and $T$.
See Appendix~\ref{app:sum_rule} for details.

\section{Results \label{sec:result}} 
\subsection{Half filling}
We first focus on the half-filled case.
Due to the chiral symmetry, the half-filling is achieved by setting $\mu = 0$. 
Figure~\ref{fig:result_HF}(a) shows $\chi_{\rm orb}$
in the parameter space $(t^\prime/t)-(t_g/t)$.
We see the remarkable parameter dependence of $\chi_{\rm orb}$.
When one of $|t^\prime|$ and $|t_g|$ is much smaller than the other,
i.e., the system is close to the pure honeycomb ladder, 
$\chi_{\rm orb}$ is negative, 
which agrees with with the result of the pure honeycomb ladder~\cite{Wakabayashi1999}.
On the other hand, when $t^\prime$ or $t_g$ are comparable to each other, 
the response becomes paramagnetic.
This is consistent with the result of the pure two-leg ladder~\cite{Roux2007}.
To gain further insight on this result, in Figs.~\ref{fig:result_HF}(b) and \ref{fig:result_HF}(c), we plot $\chi^{\rm Sea}_{\rm orb}$ and $\chi_{\rm orb}^{\rm Surf}$, respectively. 
We see that, for the whole parameter range, 
$\chi^{\rm Sea}_{\rm orb}$ is 
diamagnetic, whereas 
$\chi_{\rm orb}^{\rm Surf}$ is, as mentioned in the previous section, paramagnetic.
Thus, the sign of $\chi_{\rm orb}$ is determined by the competition between these two contributions 
with the opposite sign. 

For the pure honeycomb ladder, (i.e., $t^\prime = t$ and $t_g=0$),
we have $\chi_{\rm orb} \sim -0.13 \chi_0$
(hence $-0.13 A_0^2$ per unit cell).
The absolute value of 
the susceptibility is slightly larger than that of the assembly  of  
the $L$ independent hexagonal rings,
where the susceptibility per hexagon (at $T=0$) is
$\chi_{\rm orb} = -\frac{1}{9} A_0^2 \sim -0.11A_0^2 $; see Appendix~\ref{app:hexagon}.
This indicates 
that 
the formation of one-dimensional solid  gives rise to the additional diamagnetic contribution for the pure honeycomb ladder.

\subsection{Chemical potential dependence}
Next, we argue the $\mu$ dependence of $\chi_{\rm orb}$.
Figure~\ref{fig:result} shows the 
$\mu$ dependence 
of $\chi_{\rm orb}$. 
We see that $\chi_{\rm orb}$ is very sensitive to $\mu$. Around $\mu \sim 0$, the diamagnetic susceptibility is maximized at $\mu=0$ for $t_g=0$ (i.e., the minimal plaquette is a hexagon),
while the minima shift to the finite $\mu$ for small $|t_g|$.

Far away from half-filling, 
we see the diamagnetic response at the band top and the bottom, regardless of the hopping parameters. This is reminiscent of the Landau diamagnetism~\cite{Landau1930} for the electron gas. However, 
the analogy is not apparent since the model is one-dimensional and the Landau quantization is absent. 

Another characteristic property is the paramagnetic response for $t^\prime = -1.5$ (green lines)
for the chemical potential being at the band gap between the first and the second bands [e.g., $ -1.5 \lesssim \mu \lesssim -1.2$ for Fig.~\ref{fig:result}(c)] and the third and the fourth bands.
In the band gap, the Fermi surface term is vanishingly small for the low temperatures, hence the Fermi sea term is dominant.
Interestingly, these band gaps
are sandwiched by the divergingly large DOS coming from the top of the valence band and the bottom of the conduction bands. 
This situation is reminiscent of the effective model for the nodal line semimetal, where the orbital paramagnetism appears when $\mu$ is between two van Hove singularities~\cite{Ozaki2024}.
In that model, this paramagnetism is attributed to the interband effect,
whereas our formulation cannot specify the interband effect 
in the context of Fukuyama's  formula~\cite{Fukuyama1971,Ogata2015}
where the magnetic field is treated perturbatively.
Nevertheless, it is interesting to find similar paramagnetism in the present one-dimensional model.

For comparison, we calculate the orbital magnetic susceptibility using the Green's-function-based method\cite{GomezSantos2011, Raoux2015} with sets of model parameters used in Fig.~\ref{fig:result}.
Although this method is often used for two-dimensional systems, 
it is applicable to one-dimensional models by employing the real-space-based definition of the velocity operators.
As expected, the results are in excellent agreement with the results obtained from our straightforward method.
This coincidence confirms the applicability of the Green's-function-method to quasi-one-dimensional systems.
The details of the calculations are provided in Appendix~\ref{app:gomezsantos}.

\subsection{Current distribution in real space}
\begin{figure}[tb]
\begin{center}
\includegraphics[clip,width = \linewidth]{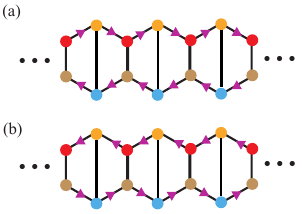}
\vspace{-10pt}
\caption{(Color online) The current distribution in the real space for $(t,t^\prime, t_g, \phi,k_{\rm B}T) = (-1,-1.5,-0.1, 0.005,0.01)$ with
(a) $\mu = 0$ (i.e., the diamagnetic case), and (b) $\mu = -1.36$ 
(i.e., the paramagnetic case).
The absolute value of the current is 
$3.2\times 10^{-4}$ for (a) and $4.3\times 10^{-4}$ for (b).
}
\label{fig:current}
\end{center}
\vspace{-10pt}
\end{figure}

To gain insight into para- and diamagnetic susceptibilities, we investigate the current distribution under the magnetic field.
We define the current operator 
at the bond between the sites $(R,\alpha)$ and $(R+\delta,\alpha^\prime)$
as  
\begin{align}
J_{(R,\alpha),(R+\delta,\alpha^\prime)}
:= -i\left[t_{(R,\alpha),(R+\delta,\alpha^\prime)}
c^{\dagger}_{R,\alpha} c_{R+ \delta, \alpha^\prime}
-(\mathrm{h.c.})
\right],
\end{align}
where $t_{(R,\alpha),(R+\delta,\alpha^\prime)}$ denotes the hopping at this bond, 
including the Peierls phase 
for the bonds with arrows in Fig.~\ref{fig:model}.
Using translational symmetry, we can calculate the 
expectation value of the current operator 
using the momentum-space representation as 
 \begin{align}
& \langle J_{(R,\alpha),(R+\delta,\alpha^\prime)} \rangle
= \frac{2}{L} \notag \\
&\times\sum_{k,\eta_1,\eta_2}\mathrm{Im}
\left\{ 
t^{\delta,\alpha,\alpha^\prime}_{k}
\left[u^{\alpha}_{k;\eta_1,\eta_2}(\phi)\right]^\ast u^{\alpha^\prime}_{k;\eta_1,\eta_2}(\phi)\right \}
f(\varepsilon_{k;\eta_1,\eta_2}(\phi))
,
 \end{align}
 with $t^{\delta,\alpha,\alpha^\prime}_{k} = t_{(R,\alpha),(R+\delta,\alpha^\prime)}e^{ik(r_{\alpha^\prime}+\delta-r_{\alpha})}$.
Figure~\ref{fig:current} shows the current distribution for small but finite $\phi$, where the purple arrows indicate the direction of the current. 
We see that no current flows on the vertical bonds, which coincides with the case of the graphene nanoribbons with zigzag edges~\cite{Wakabayashi1999}.
This is due to the periodic boundary condition;
in fact, under the open boundary condition,
finite current can flow on 
the vertical bonds as indicated in Ref.~\cite{Roux2007}.
The amount of the current are the same for the upper and the lower rows. 
As expected, comparing the diamagnetic [Fig.~\ref{fig:current}(a)] and the paramagnetic [Fig.~\ref{fig:current}(b)] cases, 
the current flows in the opposite direction to each other, resulting in the opposite direction of the current-induced magnetic moment. 

\section{Summary and discussions \label{sec:summary}}
We have investigated the orbital magnetic susceptibility 
of the honeycomb ladder.
We calculate the thermodynamics potential 
and its magnetic-field derivatives directly 
by using the single-particle eigenenergies 
as a function of the magnetic field.
Focusing on the low-temperature case, 
we have found that the response is very sensitive to the hopping parameters 
and the chemical potential.
At half-filling, the $\chi_{\rm orb}$ is diamagnetic when the system is close to the pure honeycomb ladder, 
whereas it turns to paramagnetic when it is close to the two-leg ladder.
Away from half-filling, we have found
the diamagnetism at the band tops and bottoms being reminiscent of the Landau diamagnetism,
and the large paramagnetism as the band gap sandwiched by the divergent DOS being reminiscent of the interband paramagnetism.
Additionally, we have confirmed that the orbital magnetic susceptibility obtained through our straightforward method agrees well with that calculated using the Green's-function-based approach.

We close this paper by addressing future perspectives.
From the material's point of view, 
the magnetic susceptibility measured in the experiment is fitted by the sum of the Curie-Weiss-type contribution and the Bleaney-Bowers-type contribution~\cite{Kitao2023}, 
implying that the spin part is dominant rather than the orbital part.
Besides, from the fundamental viewpoint,
it will be an interesting future problem to consider the effect of 
the electron-electron interaction and the electron-phonon coupling ~\cite{Kivelson1983,Hachmann2007,
Karakonstantakis2013,Schmitteckert2017,Dressler2018}
on the orbital magnetic susceptibility, which will be accessible by using, 
e.g., the density matrix renormalization group method~\cite{Roux2007,Tada2022}.

\begin{acknowledgement}
We thank Masao Ogata for fruitful discussions. 
This work is supported by 
JSPS KAKENHI, Grant 
No.~JP23K03243 and No.~JP21H05191.
\end{acknowledgement}

\appendix 
\section{Sum rule for $\chi_{\rm orb}$ \label{app:sum_rule}}
In this appendix,
we prove the sum rule for $M$ and $\chi_{\rm orb}$ holds for this model.
First, we consider $M$:
\begin{align}
    M&=-A_0 \sum_{\eta_1, \eta_2, k}  \frac{\partial \varepsilon_{k;\eta_1,\eta_2}(\phi)}{\partial \phi} f(\varepsilon_{k;\eta_1, \eta_2}(\phi)) \nonumber \\
    &=-A_0 \sum_{\eta_2, k}  \frac{\partial \sqrt{\Lambda_{k;\eta_2}(\phi)}}{\partial \phi}
    \left\{ f\left(\sqrt{\Lambda_{k;\eta_2}(\phi)}\right)-f\left(-\sqrt{\Lambda_{k;\eta_2}(\phi)}\right)\right\}.
\end{align}
To proceed further, it is important to notice that~\cite{ozakiogata2023}
\begin{align}
    \int_{-\infty}^\infty (f(x)-f(-x))d\mu &=  
    -\int_{-\infty}^\infty d\mu \frac{\partial}{\partial \mu}(F(x,\mu)-F(-x,\mu)) \nonumber \\
    &=\frac{1}{\beta} \int_{-\infty}^\infty d\mu \frac{\partial}{\partial \mu}
    \log \frac{1+e^{-\beta(x-\mu)}}{1+e^{-\beta(-x-\mu)}} \nonumber \\
    &=-2x, \label{eq:identity}
\end{align}
where $F(x,\mu)=-\frac{1}{\beta}\log(1+e^{-\beta(x-\mu)})$.
Then, performing $\mu$ integral with using Eq.~(\ref{eq:identity}), we obtain
\begin{align}
    \int_{-\infty}^\infty M d\mu &=
    A_0 \sum_{\eta_2, k} 2\sqrt{\Lambda_{k;\eta_2}(\phi)} 
    \frac{\partial \sqrt{\Lambda_{k;\eta_2}(\phi)}}{\partial \phi} \nonumber \\
    &=A_0 \sum_{\eta_2,k}\frac{\partial \Lambda_{k;\eta_2}(\phi)}{\partial \phi} \nonumber \\
    &=A_0 \sum_{k}\frac{\partial}{\partial \phi}a(k,\phi) \nonumber \\
    &=-A_0 \sum_{k} \frac{\partial}{\partial k}a(k,\phi) \nonumber \\
    &=0,
\end{align}
where we have also used the fact that $a(k,\phi)$ is a periodic function of $k$ in the last line.
Since this identity holds for arbitrary $\phi$, we obtain 
the sum rule for $\chi_{\rm orb}$
by performing the $\phi$-derivative,
\begin{equation}
    \int_{-\infty}^\infty \chi_{\rm orb} d\mu
   = \int_{-\infty}^\infty \frac{\partial M}{\partial \phi} d\mu
    = \frac{\partial }{\partial \phi}\left( \int_{-\infty}^\infty M d\mu\right)
    =0.
\end{equation}

\section{Orbital magnetic susceptibility for a single hexagonal ring \label{app:hexagon}}
In this appendix, we review the orbital magnetic susceptibility for a single hexagonal ring~\cite{London1937,Matsuura2016}. 
Here we set the hoppings as $t =  -1$.
In the presence of the flux $\phi$, 
the six eigenenergies are given as 
\begin{align}
\varepsilon_{n,\phi} = -2 \cos  \left( \frac{\pi}{3}n +\frac{\phi}{6}\right),
\end{align}
with $n = -2, -1,0,1,2, 3$. 
At half-filling and $T=0$, 
the states with $n=-1,0$,
and 1 are occupied for small $\phi$.
Then the orbital magnetic susceptibility at zero temperature is
\begin{align}
\chi_{\rm orb} = -A_0^2 
\sum_{n = -1,0,1}
\frac{\partial^2 \varepsilon_{n,\phi}}{\partial \phi^2}
= -\frac{1}{9} A_0^2.
\end{align}
This result is for zero temperature, but the value is almost the same for small temperatures, say, $k_{\rm B}T=0.01$ for which we have analyzed the honeycomb ladder in the main text.

\section{Comparison with the orbital magnetic susceptibility formula}
\label{app:gomezsantos}
In this appendix, we provide the details of the calculation of the orbital magnetic susceptibility using the Green's-function-based formula for lattice systems.
The orbital magnetic susceptibility is generally expressed as \cite{GomezSantos2011,Raoux2015}
\begin{align}
    \chi 
    =& k_B T \frac{e^2}{\hbar^2} \sum_{n,k} {\rm Tr} [\mathcal{G} \gamma_x \mathcal{G} \gamma_y \mathcal{G} \gamma_x \mathcal{G} \gamma_y \notag \\
    +& \frac{1}{2} (\mathcal{G}\gamma_x \mathcal{G} \gamma_y + \mathcal{G} \gamma_y \mathcal{G} \gamma_x)\mathcal{G} \gamma_{xy} ],
    \label{eq:gsformula}
\end{align}
where $\mathcal{G}=\mathcal{G}(i\omega_n)$ is the thermal Green's function, $\omega_n=(2n+1)\pi k_B T(n\in \mathbb{Z})$ is the Matsubara frequency.
In accordance with Ref.~\cite{Raoux2015}, we have defined a velocity operator in the $x$ ($y$) direction, $\gamma_x$ ($\gamma_y$), and its derivative $\gamma_{xy}$ as 
\begin{align}
    \gamma_x = \frac{\partial H}{\partial k},\qquad
    \gamma_y = -i[\hat{y},H], \qquad
    \gamma_{xy}=\frac{\partial \gamma_y}{\partial k},
\end{align}
where $\hat{y}=\sum_{j=1}^4 y_j c_{k,j}^\dagger c_{k,j}$ 
with $y_1=-y_3=1/\sqrt{3}$ and $y_4=-y_2=2/\sqrt{3}$ is the position operator in the $y$ direction.

\begin{figure}[tb]
\begin{center}
\includegraphics[clip,width = \linewidth]{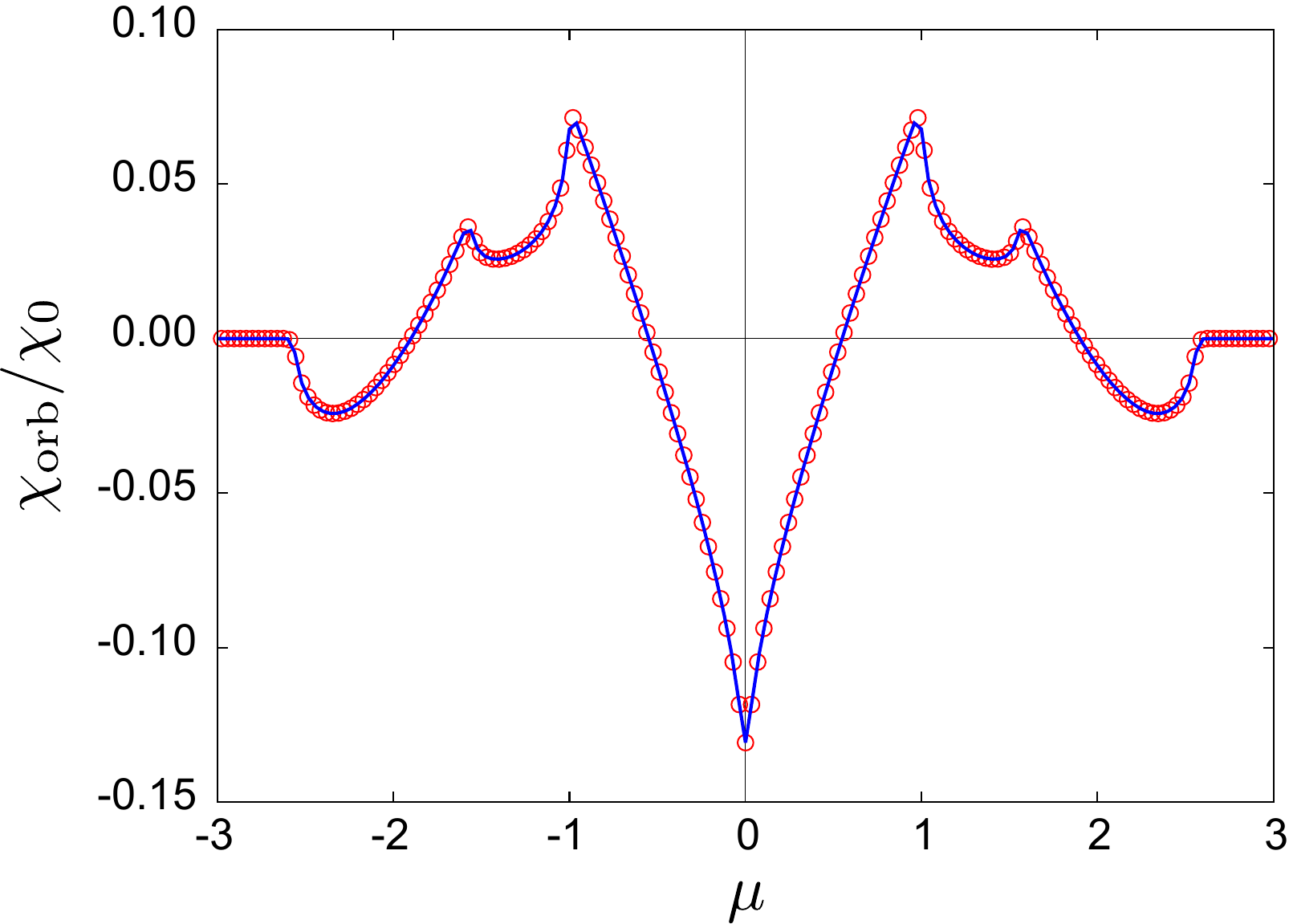}
\vspace{-10pt}
\caption{(Color online) 
Orbital magnetic susceptibility $\chi_{\rm orb}$ as functions of the chemical potential $\mu$ for the case of $(t_,t',t_g)=(-1,-1,0)$ and $k_{\rm B}T=0.01$. The solid line and open circles represent the values calculated from Eq.~\eqref{eq:chi} and from Eq.~\eqref{eq:gsformula}, respectively.
}
\label{fig:susceptcomp}
\end{center}
\vspace{-10pt}
\end{figure}

Figure~\ref{fig:susceptcomp} shows the orbital magnetic susceptibility for the case with $(t_g,t')=(0,-0.5)$ at $k_{\rm B}T=0.01$, obtained from Eq.~\eqref{eq:gsformula} and Eq.~\eqref{eq:chi}, which show excellent agreement.
Similar agreement is observed for the other parameter sets examined in Fig.~\ref{fig:result}.
The numerical evaluation of Eq.~\eqref{eq:gsformula} has been performed using the 'sparse-ir' package~\cite{shinaoka, sparsesample,wallerberger} for the Matsubara summation.
The cutoff energy for the singular value expansion $\omega_{\rm max}$ and the mesh for $k$-summation 
$N$ have been set to $\omega_{\rm max}=10|t|$ and $N=1000$, respectively. 

\bibliographystyle{jpsj}
\bibliography{polyacene}
\end{document}